\documentclass[12pt,a4paper]{article}
\usepackage{amssymb, amsthm, amsmath}
\usepackage[a4paper]{geometry}
\usepackage{amssymb,latexsym,amscd}
\usepackage{epsfig}
\usepackage{graphics,graphicx}
\usepackage{color}
\usepackage{natbib}

\definecolor{r}{rgb}{1,0,0}
\definecolor{b}{rgb}{0,0,1}

\newcommand{\tr}[1]{\textcolor{r}{#1}}
\newcommand{\tb}[1]{\textcolor{b}{#1}}

\newcommand{\beqn}{\begin{eqnarray}\begin{aligned}}
\newcommand{\eqn}{\end{aligned}\end{eqnarray}}

\theoremstyle{plain}

\def\blfootnote{\xdef\@thefnmark{}\@footnotetext} 

\theoremstyle{definition}

\title{Is the general time-reversible model bad for molecular phylogenetics?}
\author{Jeremy Sumner$^{1,\ast}$, Peter Jarvis$^{1,\dagger}$, Jes\'us Fern\'andez-S\'anchez$^{2}$,\\ Bodie Kaine$^{1}$, Michael Woodhams$^{1}$, and Barbara Holland$^{1,\mathdollar}$}

\begin{document}
\maketitle

\vspace{-1em}

\footnotesize{\noindent
$^1$School of Mathematics and Physics, University of Tasmania, Australia\\
$^2$Departament de Matem\`atica Aplicada, Universitat Polit\`ecnica de Catalunya, Spain\\
$^{\dagger}$Alexander von Humboldt Fellow, $^{\ast}$ARC Research Fellow, $ ^{\mathdollar}$ARC Future Fellow\\
\textit{keywords:} phylogenetics, model selection, General Time Reversible (GTR) model, closure\\
\textit{email:} jsumner@utas.edu.au}
\normalsize

\vspace{-3em}
\section*{}
The general time-reversible (GTR) model \citep{tavare1986} has been the workhorse of molecular phylogenetics for the last decade. 
GTR sits at the top of the \texttt{ModelTest} hierarchy of models \citep{posada1998} and, usually with the addition of invariant sites and a gamma distribution of rates across sites, is currently by far the most commonly selected model for phylogenetic inference (see Table~\ref{tab:GTRpop}).

\begin{table}[htb]
\center
\begin{tabular}{l|rrrrrr}
\hline
\multicolumn{6}{c}{Period} \\
\cline{1-7}
{\bf Phylogeny+} & 1981-85 & 1986-90 & 1991-95 & 1996-2000 & 2001-05 & 2006-10 \\
\hline
{\bf JC69}  & 29	& 127 &	692 & 1730 & 2720 & 3530 \\
{\bf K80}   & 0 &	2 &	78 & 589 & 1630 & 2920 \\
{\bf F81}   & 0 &	0 &	7 &	64 & 229 & 288 \\
{\bf HKY} & 0 &	0 & 15 & 257 & 1320 & 2920 \\
{\bf GTR}	  & 0 &	1 &	20 & 201 & 2510 & 8370 \\
{\bf GMM}   & 1 &	1 &	1 &	18 & 41 & 90 \\
\hline
\end{tabular}
\caption{Popularity of phylogenetic models of DNA substitution as measured by number of hits in Google Scholar on the search terms shown in bold (search conducted 23/08/2011).
Note that there are a small number of false positives. 
JC69 is the Jukes-Cantor model \citep{jukes1969}, K80 is the Kimura two-parameter model \citep{kimura1981}, F81 is the Felsenstein model \citep{felsenstein1981}, HKY was introduced in \citet{hasegawa1985}, and GMM is the general Markov model \citep{barry1987}.}
\label{tab:GTRpop}
\end{table}

However, a recent publication \citep{sumner2011} shows that GTR, along with several other commonly used models, has an undesirable mathematical property that may be a cause of concern for the thoughtful phylogeneticist. 
In mathematical terms, the problem is simple: matrix multiplication of two GTR substitution matrices does not return another GTR matrix. 
It is the purpose of this article to give examples that demonstrate why this deficit may pose a problem for phylogenetic analysis. 

Consider a single molecular sequence where each site evolves under the same, albeit heterogeneous, Markov process (we assume that sites evolve independently and under identical conditions). 
For time $t\!=\!0$ to $t_1$, the sequence evolves under a time-homogeneous Markov process so that substitutions are governed by a rate-matrix $Q_1$ whose $ij$th entry is the rate at which state $i$ changes to state $j$, so that the corresponding probability substitution matrix is given by $M_1\!=\!e^{Q_1t_1}$.
Then a breakpoint occurs, and 
over time $t_2$ the sequence again evolves under a time-homogeneous Markov process but with a different rate-matrix $Q_2$ governing the substitution rates, so that the corresponding substitution matrix for this time period is $M_2\!=\!e^{Q_2t_2}$. 
As a consequence of the Markov assumption for the overall process, the substitution matrix that describes the probability of substitutions between time $t\!=\!0$ and $t\!=\!t_1+t_2$ can be expressed as the matrix product $\widehat{M} = M_1M_2$.
So far, so good, but then two questions naturally arise: 
\begin{enumerate}
\item[i.] Is there a single rate-matrix $\widehat{Q}$ that can be used to describe the process from the time $t\!=\!0$ to $t\!=\!t_1+t_2$ as a \emph{homogeneous} Markov process with $e^{\widehat{Q}(t_1+t_2)}=\widehat{M}$? 
\item[ii.] If $Q_1$ and $Q_2$ are in a particular model class, will $\widehat{Q}$ necessarily belong to the same class? 
\end{enumerate}
\citet{sumner2011} show that the answer to the first question is ``yes'' (by confirming that the general Markov model is closed under matrix multiplication), and that the answer to the second question is ``sometimes'' depending on the class of model considered.
Crucially, for the case of the GTR model class they show that the answer to the second question is ``no''.

Why does this lack of closure under matrix multiplication for the GTR model matter for phylogenetics? 
In almost all standard phylogenetic studies a model is used that assumes a homogeneous Markov process generated the molecular data.
This is implemented by taking a fixed rate-matrix to apply at all times and on all lineages of the evolutionary tree, and helps to maximize statistical power by reducing the number of free parameters present and thus keeping the corresponding estimation variance to a minimum. 
However, the thoughtful phylogeneticist does not necessarily believe that the truth of the matter is that the random process remained fixed through time and across the lineages of the evolutionary tree.
Rather, biological reality is likely to consist of some form of time-dependent \emph{and} lineage-specific evolution governed by varying substitution rates. 
Under the assumption of a Markov process, this corresponds mathematically to the rate-matrix changing across the evolutionary tree.
For example, it has long been known that base composition is not fixed throughout evolutionary history \citep{foster1997,chang2000,lockhart1992,phillips2004,tarrio2001}, and that there is evidence that the Markov process varies across the evolutionary tree \citep{herbeck2005,ota2003}.


%
%
So, it is apparent that the thoughtful phylogeneticist already thinks of her model as a statistical pragmatist's average of the true heterogeneous process.
In this light, multiplicatively closed models are then simply those where it is possible to assume a homogeneous model and ``average out'' these effects in a mathematically consistent way. 

Before discussing how the non-closure of GTR may affect phylogenetic estimation, we think it is worth reflecting on how GTR made its way to the top of the \texttt{ModelTest} hierarchy.
Reversibility of a continuous-time Markov chain $X(t)$ on some time interval $t\in\left[0,T\right]$ arises by considering the ``time-reversed process'' $Y(t)=X(T-t)$, and demanding that the probability of observing $X(T)\!=\!j$ given $X(0)\!=\!i$ is identical to the probability of observing $Y(T)\!=\!i$ given $Y(0)\!=\!j$. 
In this case, it is easy to show that the rate-matrix for the time-reversed process $Y(t)$ is exactly the same as that for the original process, and the process is said to be ``time-reversible'' or simply ``reversible''.
Importantly, time-reversibility has long been held out as important for phylogenetic analysis because implementation of Felsenstein's ``pruning'' algorithm for calculating likelihoods \citep{felsenstein1981} relies on the so-called ``pulley-principle'' and the process being reversible.
However, recent authors have developed likelihood algorithms for non-reversible processes \citep{boussau2006,oscamou2008,sumner2010b}, and, besides, we argue that algorithmic convenience is not the only criterion that the thoughtful phylogeneticist may wish to take into account when choosing appropriate models.

In case the concerns outlined above are seen to be somewhat philosophical,
we now show that lack of multiplicative closure of a model can result in systematic biases for phylogenetic inference 
If we use a model where the true heterogeneous process cannot be exactly represented homogeneously under the same model, then it is possible that some error in estimation of substitution rates and/or speciation times will occur. 
A computer simulation was conducted to test how severe this misestimation could be for the simple situation of a single sequence evolving under the GTR model with a breakpoint where the substitution process was allowed to change.
As per the situation described above, GTR rate-matrices $Q_1$ and $Q_2$ applied over times $t_1$ and $t_2$ respectively, giving the substitution matrices, $M_1=e^{Q_1t_1}$ and $M_2=e^{Q_2t_2}$, and $\widehat{M}=M_1M_2$ represents the transition matrix for the entire edge.
We normalized the rate-matrices $Q_1$ and $Q_2$ such that the sum of the (off-diagonal) rates was equal to 1, this allows edge lengths to
be compared in a consistent way.
We then sought to find a (similarly normalized) GTR class rate-matrix $\bar{Q}$ and time $\bar{t}$, such that the ``distance'' from $e^{\bar{Q}\bar{t}}$ to $\widehat{M}$ was minimized.
For this purpose we used $d(M,N)=(\sum_{i \neq j} (\left[M\right]_{ij}-\left[N\right]_{ij})^2)^{1/2}$, as the distance between any two substitution matrices $M$ and $N$ (where it should be noted that the constraint $i\neq j$ in the summation is intentional; reflecting that we are effectively calculating the Euclidean distance between substition matrices as embedded in the ambient vector space of GMM substitution matrices).
We also used the same distance to determine the distance between the two rate matrices $Q_1$ and $Q_2$, to give a measure of how ``heterogeneous'' the process was in each case. 
In each iteration of the simulation we select two rate-matrices, $Q_1$ and $Q_2$, from the GTR model class by sampling substitution rates randomly in such a way as to ensure a reasonably consistent spread between almost homogeneous and highly heterogeneous. 
All calculations were performed in the statistical computing package \texttt{R} \citep{Rproject} and its in-built optimisation routine ``optim''.

\begin{figure}[htb]
\begin{center}
\includegraphics[scale=0.45]{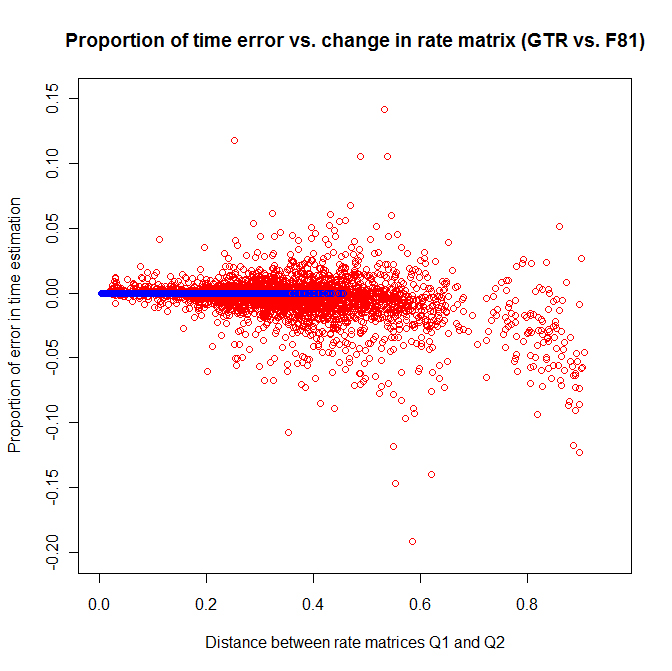}
\end{center}
\caption{Proportion of error in the time estimate for the homogeneous representation of a truly heterogeneous situation increases significantly as the amount of change in rates increases at the breakpoint.
\tb{BLUE} points show results for F81 (a closed model), and \tr{RED} points show results for GTR.
}
\label{fig:GTRF81}
\end{figure}

To assist in comparison, we repeated the above simulation for the known closed model F81 (see \citet{sumner2011} for details).
F81 was chosen as, despite it being guaranteed that $\widehat{Q}$ will also belong to the F81 model class (and hence our routine should find $\bar{Q}=\widehat{Q}$), it is also \emph{not} the case that $\widehat{Q}$ is simply given by the weighted average of $Q_1$ and $Q_2$ (which should be compared to other closed models, such as JC69 and K80, where $\widehat{Q}$ is given by the weighted mean).
For this reason, F81 was deemed as a fair counterpoint to the GTR model.
For all simulations, $t_1$ and $t_2$ were chosen to be $0.5$, such that $t_1+t_2\!=\!1$, and $\bar{t}$ was compared to the actual time proportionally. 
Figure~\ref{fig:GTRF81} shows the level of error in time estimation for both the closed F81 model and the unclosed GTR model.

\begin{figure}[tb]
\begin{center}
\includegraphics[scale=0.45]{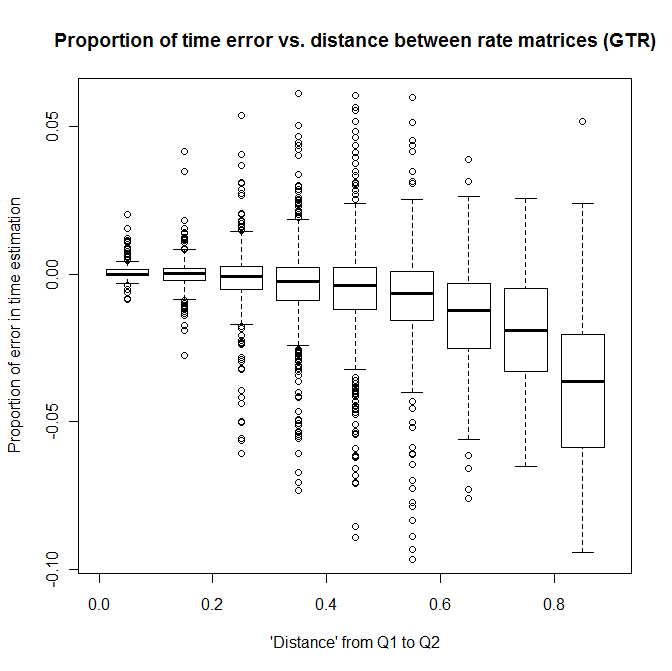}
\end{center}
\caption{Boxplots of the GTR model showing the time error increasing as the distance between the input parameters at a breakpoint increases. There appears to be a tendency towards underestimation of edge lengths.}
\label{fig:GTRboxplots}
\end{figure}

The results in Figure~\ref{fig:GTRF81} suggest that the F81 model correctly estimates the time parameter to within the tolerance of the optimisation routine.
However, for GTR we found up to a 19\% under-estimation and 14\% over-estimation of the time parameter, 
although this occurred in cases where the breakpoint introduced a large change in the rate-matrix.
The magnitude of the error in time estimation increases as the distance between the two input rate-matrices increases.
Another trend in the GTR data is that the points tend to underestimate more than overestimate as the input parameter distance increases.
Figure~\ref{fig:GTRboxplots} shows this by redisplaying the data from figure~\ref{fig:GTRF81} using a boxplot for each increment of $0.1$ in the distance between the two input rate-matrices. 
Finding a theoretical explanation for this tendancy for underestimation is currently an open problem.

Having decided that lack of closure is a problem for phylogenetic models, the obvious question that arises is what are the closed models? 
These ``good'' models are not yet fully known, and \citet{sumner2011} discuss that in general this is quite a difficult and subtle mathematical problem that crosses the boundaries of Markov chain and Lie group theory.
However, they do give a method that shows how to generate a complete list of models that have a certain invariance properties under permutations of nucleotide states.
In particular they give a complete hierarchy of closed models with 4-way nucleotide symmetry; that is, models that do not prefer any particular groupings of nucleotides, or to put it another way, models where for any relabelling (permutation) of the states it is possible to find a relabelling (permutation) of the substitution parameters such that the model is unchanged.
The hierarchy they present contains GMM (the general Markov model), K3ST, F81 and JC, but includes a newly-discovered 6-parameter closed model dubbed ``K3ST+F81'', as it combines features of both these models. 
Work on cataloguing the ``good'' models for other symmetries is underway, with the method given in \citet{sumner2011} being applied to generate the list of closed models that have the transition/transversion substition symmetry.
For example, K2P is an example of a model with this symmetry that is known to be closed, whereas HKY has the same symmetry but is not closed.

As a final example, we compared the performance of the closed model K3ST+F81 to GTR, SYM (the GTR model with uniform base distribution), and HKY.
These initial comparisons of the K3ST+F81 model to models which are not multiplicatively closed show that it tends to give sometimes better and sometimes worse likelihoods.
For five different datasets a tree topology was generated using neighbour-joining and then, fixing this tree topology, we performed likelihood under each model, fitting rate substitutions and edge lengths.
The likelihood values for the various models and datasets are presented in  Table~\ref{tab:loglikes}.
With the caveat that it is not necessarily statistically meaningful to compare likelihood values for these different models 
(especially given that the models are not nested), it can be seen from Table~\ref{tab:loglikes} that GTR consistently outperforms the other models. 
The differences in log-likelihood cannot be solely explained by its increased number of free parameters.
For the models with comparable number of parameters we found that sometimes the best fit comes from our closed model and sometimes from a non-closed model.
What conclusions can be drawn from this study? 
Our overall point is that it is not prudent to only look for models that fit the data ``best'' in a likelihood sense, 
but to also look for models that can be given a satisfactory \emph{interpretation.}
We claim that, particularly in cases where a heterogeneous process is suspected, priority should be given to a closed model even when a non-closed model gives a better likelihood.

\begin{table}[tb]
\center
\begin{tabular}{|c|ccccc|}
\hline 
\textbf{Model} &   Human    &   Acorus        &  Cormorants   & Yeast  &  Fish \\
\hline
 \textbf{F+K} (5)   	 &  {\it -1557.76} & {\it-451396.98} & {\it -7014.46} & {\it -710065.35} & {\it -14534.49} \\
 \textbf{HKY} (4)      &          -0.72  &         355.55  &         14.38  &          365.03  &           8.50  \\
 \textbf{SYM} (5)      &          -0.45  &       -1990.37  &         -3.11  &        -3695.09  &           5.70  \\
 \textbf{GTR} (8)      &           4.14  &         862.60  &         40.76  &         2409.14  &          38.86  \\
\hline
\end{tabular}

\caption{Log-likelihoods of various phylogenetic models on sample data sets. 
The number of free parameters for each model is given in parentheses (with the count reduced by one after overall scaling is taken into account).
The values for the F81+K3ST model (shortened here to ``F+K'') are the actual log likelihoods, the other values are relative to this (with positive values indicating a larger likelihood). 
The data sets are: human, with 53 taxa, 202 sites, of mitochondrial genes \citep{ingham2000}; Acorus, with 15 taxa, 89436 sites, from the chloroplast genomes \citep{goremykin2005}; Cormorants, with 33 taxa, 1141 sites, from a mix of mitochondrial and nuclear genes \citep{holland2010}; Yeast, with 8 taxa, 127026 sites, from mainly nuclear genes \citep{rokas2003}; Fish, with 11 taxa, 2178 sites, from nuclear genes \citep{zakon2006}. 
Note that differences in log-likelihood greater than $k$, where $k$ is the difference in the number of parameters, would be considered significant under the Akaike Information Criterion.}

\label{tab:loglikes}
\end{table}

To summarize, we argue that lack of closure constitutes a serious problem for the use of the GTR model in phylogenetics, as it means that taking an average of inhomogeneous processes -- which is almost certainly the underlying biological reality in many circumstances --
is impossible to do in an accurate way. 
Further research is required to find credible closed alternatives to GTR that offer similar ability to fit phylogenetic data.
It will also be important 
to determine closure for cases where inhomogeneous models are used explicitly,
for example in codon models \citep{yang1998}, which are used to test for positive selection, 
and, for instance, the work of \citet{hamady2006} which uses detectable shifts in inferred rate-matices to infer if genes have been horizontally transferred.



\bibliographystyle{jtbnew}
\bibliography{masterGTRnote}

\begin{thebibliography}{26}
\expandafter\ifx\csname natexlab\endcsname\relax\def\natexlab#1{#1}\fi
\expandafter\ifx\csname url\endcsname\relax
  \def\url#1{\texttt{#1}}\fi
\expandafter\ifx\csname urlprefix\endcsname\relax\def\urlprefix{URL }\fi
\providecommand{\selectlanguage}[1]{\relax}

\bibitem[{Barry \& Hartigan(1987)}]{barry1987}
\textsc{Barry, D. \& Hartigan, J.~A.} (1987).
\newblock Asynchronous distance between homologous {DNA} sequences.
\newblock \emph{Biometrics} \textbf{43}, 261--276.

\bibitem[{Boussau \& Gouy(2006)}]{boussau2006}
\textsc{Boussau, B. \& Gouy, M.} (2006).
\newblock Efficient likelihood computations with nonreversible models of
  evolution.
\newblock \emph{Syst. Biol.} \textbf{55}, 756--768.

\bibitem[{Chang \& Campbell(2000)}]{chang2000}
\textsc{Chang, B. S.~W. \& Campbell, D.~L.} (2000).
\newblock Bias in phylogenetic reconstruction of vertebrate rhodopsin
  sequences.
\newblock \emph{Mol. Biol. Evo.} \textbf{17}, 1220--1231.

\bibitem[{Felsenstein(1981)}]{felsenstein1981}
\textsc{Felsenstein, J.} (1981).
\newblock Evolutionary trees from {DNA} sequences: a maximum likelihood
  approach.
\newblock \emph{J. Mol. Evol.} \textbf{17}, 368--376.

\bibitem[{Foster \emph{et~al.}(1997)Foster, Jermiin \& 44:282–288}]{foster1997}
\textsc{Foster, P.~G., Jermiin, L.~S. \& 44:282–288, D. A. H. J. M. E.~.}
  (1997).
\newblock Nucleotide composition bias affects amino acid content in proteins
  coded by animal mitochondria.
\newblock \emph{J. Mol. Evol.} \textbf{44}, 282--288.

\bibitem[{Goremykin \emph{et~al.}(2005)Goremykin, Holland, Hirsch-Ernst \&
  Hellwig}]{goremykin2005}
\textsc{Goremykin, V.~V., Holland, B., Hirsch-Ernst, K.~I. \& Hellwig, F.~H.}
  (2005).
\newblock Analysis of \emph{Acorus calamus} chloroplast genome and its
  phylogenetic implications.
\newblock \emph{Mol. Biol. Evol.} \textbf{22}, 1813–--1822.

\bibitem[{Hamady \emph{et~al.}(2006)Hamady, Betterton \& Knight}]{hamady2006}
\textsc{Hamady, M., Betterton, M.~D. \& Knight, R.} (2006).
\newblock Using the nucleotide substitution rate matrix to detect horizontal
  gene transfer.
\newblock \emph{BMC Bioinformatics} \textbf{7}, 476.

\bibitem[{Hasegawa \emph{et~al.}(1985)Hasegawa, Kishino \& Yano}]{hasegawa1985}
\textsc{Hasegawa, M., Kishino, H. \& Yano, T.} (1985).
\newblock Dating of human-ape splitting by a molecular clock of mitochondrial
  {DNA}.
\newblock \emph{J. Mol. Evol.} \textbf{22}, 160--174.

\bibitem[{Herbeck \emph{et~al.}(2005)Herbeck, Degnan \&
  Wernegreen}]{herbeck2005}
\textsc{Herbeck, J.~T., Degnan, P.~H. \& Wernegreen, J.~J.} (2005).
\newblock Nonhomogeneous model of sequence evolution indicates independent
  origins of primary endosymbionts within the enterobacteriales
  (c-proteobacteria).
\newblock \emph{Mol. Biol. Evol.} \textbf{22}, 520–--532.

\bibitem[{Holland \emph{et~al.}(2010)Holland, Spencer, Worthy \&
  Kennedy}]{holland2010}
\textsc{Holland, B.~R., Spencer, H.~G., Worthy, T.~H. \& Kennedy, M.} (2010).
\newblock Identifying cliques of convergent characters: concerted evolution in
  the cormorants and shags.
\newblock \emph{Syst. Biol.} \textbf{59}, 433--445.

\bibitem[{Ingman \emph{et~al.}(2000)Ingman, Kaessmann, Paabo \&
  Gyllenstern}]{ingham2000}
\textsc{Ingman, M., Kaessmann, H., Paabo, S. \& Gyllenstern, U.} (2000).
\newblock Mitochondrial genome variation and the origin of modern humans.
\newblock \emph{Nature} \textbf{408}, 708--713.

\bibitem[{Jukes \& Cantor(1969)}]{jukes1969}
\textsc{Jukes, T.~H. \& Cantor, C.~R.} (1969).
\newblock \emph{Mammalian protein metabolism.}, chap. Evolution of Protein
  Molecules.
\newblock New York: Academic Press, pp. 21--132.

\bibitem[{Kimura(1981)}]{kimura1981}
\textsc{Kimura, M.} (1981).
\newblock Estimation of evolutionary distances between homologous nucleotide
  sequences.
\newblock \emph{Proc. Natl. Acad. Sci.} \textbf{78}, 1454–--1458.

\bibitem[{Lockhart \emph{et~al.}(1992)Lockhart, Howe, Bryant, Beanland \&
  Larkum}]{lockhart1992}
\textsc{Lockhart, P., Howe, C., Bryant, D., Beanland, T. \& Larkum, A.} (1992).
\newblock Substitutional bias confounds inference of cyanelle origins from
  sequence data.
\newblock \emph{J. Mol. Evol.} \textbf{34}, 153--162.

\bibitem[{Oscamou \emph{et~al.}(2008)Oscamou, McDonald, Yap, Huttley, Lladser
  \& Knight}]{oscamou2008}
\textsc{Oscamou, M., McDonald, D., Yap, V.~B., Huttley, G.~A., Lladser, M.~E.
  \& Knight, R.} (2008).
\newblock Comparison of methods for estimating the nucleotide substitution
  matrix.
\newblock \emph{BMC Bioinformatics} \textbf{9}, 511.

\bibitem[{Ota \& Penny(2003)}]{ota2003}
\textsc{Ota, R. \& Penny, D.} (2003).
\newblock Estimating changes in mutational mechanisms of evolution.
\newblock \emph{J. Mol. Evol.} \textbf{57}, S233–--S240.

\bibitem[{Phillips \emph{et~al.}(2004)Phillips, Delsuc \& Penny}]{phillips2004}
\textsc{Phillips, M.~J., Delsuc, F. \& Penny, D.} (2004).
\newblock Genome-scale phylogeny and the detection of systematic biases.
\newblock \emph{Mol. Biol. and Evol.} \textbf{21}, 1455--1458.

\bibitem[{Posada \& Crandall(1998)}]{posada1998}
\textsc{Posada, D. \& Crandall, K.~A.} (1998).
\newblock Modeltest: testing the model of {DNA} substitution.
\newblock \emph{Bioinformatics} \textbf{14}, 817--818.

\bibitem[{{R Development Core Team}(2006)}]{Rproject}
\textsc{{R Development Core Team}} (2006).
\newblock \emph{R: A Language and Environment for Statistical Computing}.
\newblock R Foundation for Statistical Computing, Vienna, Austria.

\bibitem[{Rokas \emph{et~al.}(2003)Rokas, amd N.~King \& Carroll}]{rokas2003}
\textsc{Rokas, A., amd N.~King, B. L.~W. \& Carroll, S.~B.} (2003).
\newblock Genome-scale approaches to resolving incongruence in molecular
  phylogenies.
\newblock \emph{Nature} \textbf{425}, 798--804.

\bibitem[{Sumner \& Charleston(2010)}]{sumner2010b}
\textsc{Sumner, J. \& Charleston, M.} (2010).
\newblock Phylogenetic estimation with partial likelihood tensors.
\newblock \emph{J Theor Biol.} \textbf{262}, 413--24.

\bibitem[{Sumner \emph{et~al.}(2011)Sumner, Fern\'andez-S\'anchez \&
  Jarvis}]{sumner2011}
\textsc{Sumner, J., Fern\'andez-S\'anchez, J. \& Jarvis, P.} (2011).
\newblock Lie {M}arkov models.
\newblock \emph{in submission, ArXiv e-prints: 1105.4680} , 32 pp.

\bibitem[{Tarr\'io \emph{et~al.}(2001)Tarr\'io, Rodr\'iguez-Trelles \&
  Ayala}]{tarrio2001}
\textsc{Tarr\'io, R., Rodr\'iguez-Trelles, F. \& Ayala, F.~J.} (2001).
\newblock Shared nucleotide composition biases among species and their impact
  on phylogenetic reconstructions of the drosophilidae.
\newblock \emph{Mol. Biol. Evol.} \textbf{18}, 1464--1473.

\bibitem[{Tavar\'e(1986)}]{tavare1986}
\textsc{Tavar\'e, S.} (1986).
\newblock Some probabilistic and statistical problems in the analysis of dna
  sequences.
\newblock \emph{Lectures on Mathematics in the Life Sciences (American
  Mathematical Society)} \textbf{17}, 57--86.

\bibitem[{Yang(1998)}]{yang1998}
\textsc{Yang, Z.} (1998).
\newblock Likelihood ratio tests for detecting positive selection and
  application to primate lysozyme evolution.
\newblock \emph{Mol. Biol. Evo.} \textbf{15}, 568--573.

\bibitem[{Zakon \emph{et~al.}(2006)Zakon, Lu, Zwickl \& Hillis}]{zakon2006}
\textsc{Zakon, H.~H., Lu, Y., Zwickl, D.~J. \& Hillis, D.~M.} (2006).
\newblock Sodium channel genes and the evolution of diversity in communication
  signals of electric fishes: Convergent molecular evolution.
\newblock \emph{Proc. Natl. Acad. Sci. USA} \textbf{103}, 3675--80.

\end{thebibliography}

\end{document}